\documentclass[aps,prb,reprint,superscriptaddress,showpacs]{revtex4-1}

\usepackage{graphicx,color}
\usepackage{amsmath,amssymb}

\begin{document}


\title{Photonic band structure and effective medium properties of doubly-resonant 
core-shell metallo-dielectric nanowire arrays: low-loss, isotropic optical negative-index behavior}


\author{D. R. Abujetas}
\email[]{diego.romero@iem.cfmac.csic.es}
\affiliation{Instituto de Estructura de la Materia (IEM-CSIC),
Consejo Superior de Investigaciones Cient{\'\i}ficas, Serrano 121,
28006 Madrid, Spain}
\author{R. Paniagua-Dom\'{\i}nguez}
\altaffiliation{Present address: Data Storage Institute, Agency for Science, Technology and Research, 117608 Singapore.}
\affiliation{Instituto de Estructura de la Materia (IEM-CSIC),
Consejo Superior de Investigaciones Cient{\'\i}ficas, Serrano 121,
28006 Madrid, Spain}
\author{M. Nieto-Vesperinas}
\affiliation{Instituto de Ciencia de Materiales de Madrid (ICMM-CSIC),
Consejo Superior de Investigaciones Cient{\'\i}ficas,
Campus de Cantoblanco, 28049 Madrid, Spain}
\author{J. A. S\'anchez-Gil}
\email[]{j.sanchez@csic.es}
\affiliation{Instituto de Estructura de la Materia (IEM-CSIC),
Consejo Superior de Investigaciones Cient{\'\i}ficas, Serrano 121,
28006 Madrid, Spain}


\date{\today}

\begin{abstract}

We investigate theoretically and numerically the photonic band structure in the optical domain of an array of core-shell 
metal-semiconductor nanowires.  Corresponding negative-index photonic bands are calculated, showing isotropic equifrequency 
surfaces. The effective (negative) electric permittivity and magnetic permeability, retrieved from S-parameters, 
are used to  compare the performance of such nanowire arrays with homogeneous media in canonical examples, such as refraction 
through a prism and flat-lens focusing. Very good agreement is found, indeed confirming the effective 
medium behavior of the nanowire array as a low-loss, isotropic (2D) and bulk, optical negative index metamaterial. Indeed, 
disorder is introduced to further stress its robustness.

\end{abstract}


\maketitle

\section{Introduction}

Experiments in composite systems looking for a negative index of refraction (both permittivity and 
permeability being negative) at gigahertz frequencies\cite{Shelby2001} were the first of this kind aiming at controlling the flow of electromagnetic waves; indeed, some fascinating properties have been since explored based on such composites \cite{Veselago1968,Pendry2000,Garcia2002,Cubukcu2003,Grbic2004,Ramakrishna2005,%
Leonhardt2006,Pendry2006,MMSBook,Zhang2008}.
However, to scale down these properties, not found in natural media, from the gigahertz regime to the optical domain, has posed a major challenge \cite{Shalaev2007,Wegener2011,Xu2013}, especially with regard to the negative magnetic permeability  \cite{Zhou2005,Alu2006} (negative electric response is trivial in the optical domain with metals). Despite some drawbacks of the original design, as for instance anisotropy, the initial philosophy to achieve this goal was to miniaturize the original ones, namely,  split-ring resonators, magneto-dielectric elements, and related \cite{MMSBook,Shalaev2007,Wegener2011}. Nonetheless, the saturation of the magnetic response at high frequencies restricts this approach to the far infrared \cite{Zhou2005}.  Other designs have been proposed that somehow mimic their behavior \cite{Shalaev2007,Zhang2006,Yao2008,Valentine2008,Burgos2010}, one of the most widely employed being metallic fishnets \cite{GarciaMeca2011}. The emergence of ohmic losses in metals in the optical domain hinders also their performance.

To overcome this problem, various attempts have been made  by exploiting magnetic resonances occurring in structures 
made of high-permittivity materials \cite{Antonoyiannakis1997,OBrien2002,Wheeler2005,Peng2007,Schuller2007,Popa2008,Vynck2009,Zhao2009,Zhou2009,Kuester2011}, some of them combined  with secondary plasmonic structures providing the electrical response  \cite{Yannopapas2007,Kussow2008,Jelinek2010,Kang2011}. 
Very recently, semiconductor nanostructures exhibiting magnetic resonances have been proposed 
as possible non-absorbing metamaterial elements as well as emitters with unusual properties 
\cite{OE2011,MNV2011,Kuznetsov2012,Fu2013,Person2013}. 
Actually, simple arrangements based on this idea have been proposed such as core-shell 3D and 2D metallo-dielectric nanostructures, 
respectively, nanospheres and nanowires, geometrically tuned so that their two  first order Mie resonances (magnetic and electric) 
spectrally overlap, which might play the role of doubly-resonant meta-atoms \cite{NJP2011,SR2012,Ramadurgam2014}, directional scatterers \cite{Liu2012,Liu2014}, and transparent nanowires \cite{OE13,Liu2013g,Liu2015}. Beside their 
robustness and isotropy stemming from their  highly symmetric, single-structure meta-atoms, a theoretical figure of merit $\left(f.o.m.=-
\Re(n_{\mathrm{eff}})/\Im(n_{\mathrm{eff}})\right)$ of the order of $f.o.m. \sim 200$, has been reported in the case of core-shell nanowire 
(NW) arrays \cite{SR2012}. Incidentally, NW arrays have been proposed in turn as hyperbolic metamaterials \cite{Poddubny2013}.

Nevertheless, a full description of their wave propagation properties must be done to determine whether their negative refraction behavior 
is a photonic-crystal diffraction effect that depends on the geometry sample, or a fully isotropic effective negative-index behavior
(which we will referred to as negative-index metamaterial, NIM). In fact, it has been pointed out in Ref.~\onlinecite{Valdivia2012} that the 
constitutive parameters retrieved through effective medium theories in other dielectric-rod ordered arrays do not reproduce the negative 
refraction observations in those composites, being mostly due to photonic crystal diffraction 
\cite{Foteinopoulou2003a,Garcia-Pomar2005,Decoopman2006}.  
In this work, we thoroughly investigate the photonic band structure of an array of core-shell metal-semiconductor NWs, which was shown
in Ref.~\cite{SR2012} through parameter retrieval  to exhibit low-loss, negative index behavior. This allows us to rigorously demonstrate 
(through equifrequency surfaces) and improve (fine tuning core-shell geometrical parameters) its highly isotropic character. Indeed, 
taking advantage of high isotropy and low losses, flat-lens focusing is presented through thick slabs  with subwavelength resolution and
dipole position independence, in good agreement with the expected behavior for a homogeneous slab with effective parameters as 
retrieved for the NW array. Limitations are also investigated with regard to image resolution, photonic-crystal surface 
termination, and disorder. 

This paper is organized as follows.
Corresponding negative-index photonic bands and equifrequency surfaces are calculated in Sec. II, confirming the effective medium 
behavior. In Sec. III,  the performance of such NW arrays in the canonical example of flat-lens focusing  through a thick slab,
as compared with a homogeneous slab with the retrieved electric permittivity and magnetic permeability. This is also done for
plane wave refraction through a prism in Sec. IV,  introducing in turn disorder to support the robustness of the NW array as a low-loss 
optical NIM. Finally, Section V contains our concluding remarks.

\section{Photonic band structure}

\begin{figure}
\includegraphics[width=\columnwidth]{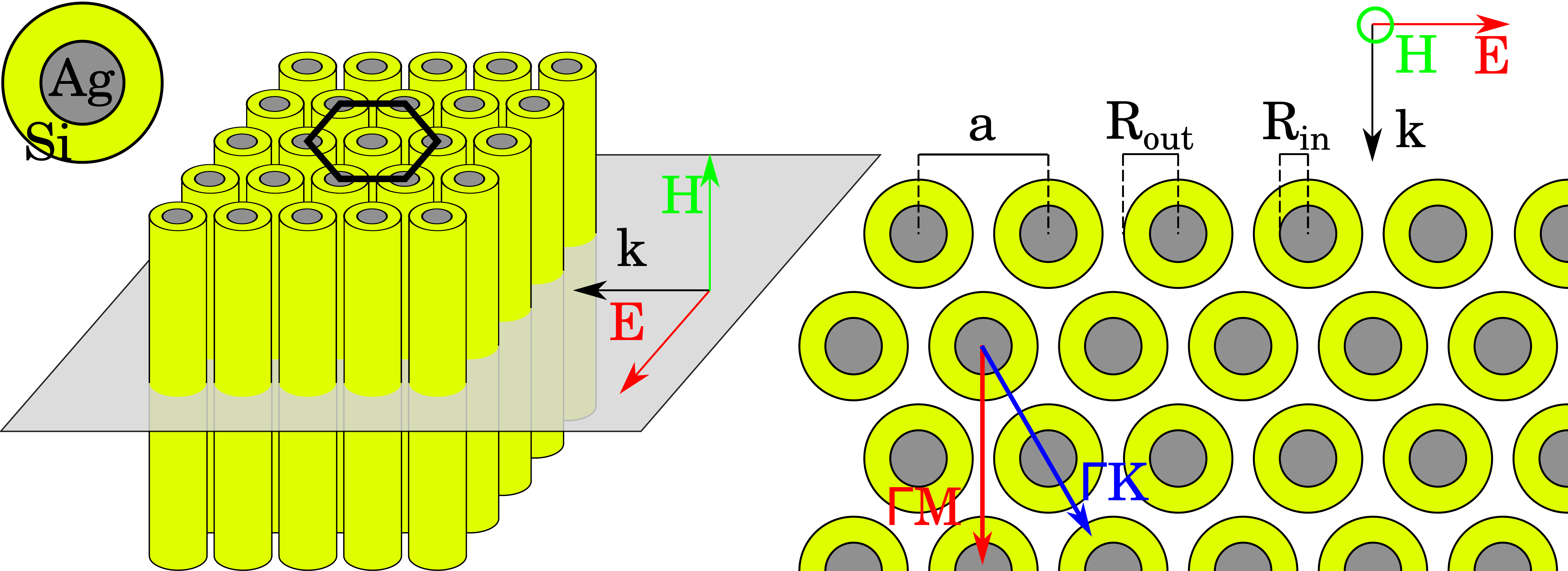}
\caption{(Color online) Schematic of the photonic crystal configuration under study, an hexagonal array of core-shell nanowires, including the relevant geometrical parameters. Only propagation in the plane perpendicular to the (infinitely long) NW axis will be considered with transverse electric polarization.}
\label{Fig-CSNWarray}
\end{figure}
\begin{figure}
\includegraphics[width=\columnwidth]{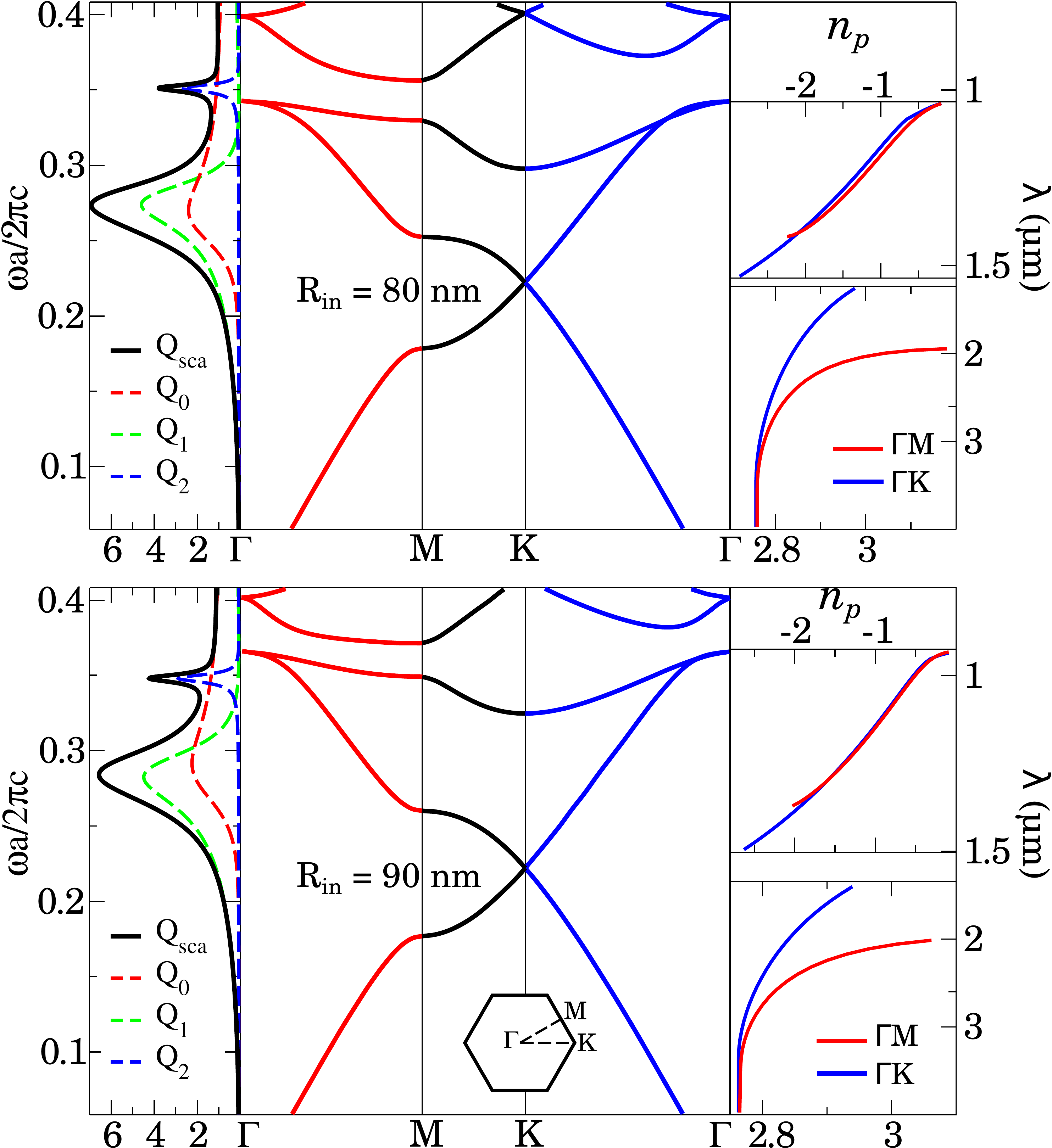}
\caption{(Color online) In-plane photonic band structure (center) of an hexagonal array ($a=$350 nm) of Ag@Si core-shell nanowires with external radius $R_{\mathrm{out}}=$ 170 nm and internal radii: (top)  $R_{\mathrm{in}}=$ 80 nm and (bottom) $R_{\mathrm{in}}=$ 90 nm. In both plots: 
(left) The Mie scattering efficiency for an isolated nanowire is also shown  (black curve), including separately the contributions from the first three multipolar terms (red, green, and blue curves); (right) phase index $n_p$ corresponding to the first two bands along the main $\Gamma$M (red curves) and $\Gamma$K (blue curves) directions.
}
\label{Fig-PhBand}
\end{figure}

First of all, we characterize the electromagnetic wave propagation in the 2D  periodic composite medium consisting of  core-shell metal-dielectric (infinitely long)  NWs in a regular hexagonal array (see Fig.~\ref{Fig-CSNWarray}). For this purpose, we plot in Fig.~\ref{Fig-PhBand}  the 2D photonic band structure for the maximum symmetry directions for TE-polarized waves (magnetic field along the NW axis) of an hexagonal array with a lattice parameter of $a=$ 350 nm made of Ag@Si core-shell NWs with external radius $R_{\mathrm{out}}=$ 170 nm, and inner radii $R_{\mathrm{in}}=80,90$ nm.  Details on the FEM numerical calculations of the photonic band structure are given in the appendix. Also included in Fig.~\ref{Fig-PhBand} is the total Mie scattering efficiency of a single Ag@Si NW, obtained from Mie formulae for infinitely long core-shell cylinders \cite{Aden1951}, explicitly plotting the contributions of the first three terms ($Q_0,Q_1,Q_2$) stemming, respectively, from the zero-order magnetic, first-order (dipolar) electric, and second-order (quadrupolar) electric terms in the Mie multipolar expansion. The first one is due to a strong circulation of the electric displacement field inside the dielectric shell, so the position of the resonance mostly depends on the shell external radius. The $Q_1$ contribution corresponds to the localized surface plasmon resonance of the metal core. For obvious reasons, i.e. NIM behavior \cite{SR2012}, we choose the appropriate geometrical parameters so that both (electric and magnetic) resonances spectrally overlap, as clearly seen in Fig.~\ref{Fig-PhBand}; indeed, another inner radius $R_{\mathrm{in}}=90$ nm has been considered that slightly improves such NIM behavior, as will be shown below. In addition, the lattice constant is such that the high filling fraction condition needed to achieve doubly negative effective parameters is fulfilled \cite{SR2012}; the lattice choice, hexagonal, favors this close packing and  in turn isotropy.

The photonic bands present a partial gap in the $\Gamma$M direction in the first band and a full gap between the third and fourth bands.  Besides, the slope of the second one is negative, and corresponds to a mode propagation with a negative phase velocity, as evidenced in the right graphs in Fig.~\ref{Fig-PhBand}, where the phase index $n_p$ corresponding to the first two bands along the main $\Gamma$M and $\Gamma$K directions is plotted. Furthermore, this band lies on the spectral region where the magnetic and electric resonances overlap, making it specially relevant to elucidate the effective NIM behavior stemming from strong individual electromagnetic responses of the isolated NWs. Thus let us study in detail this negative-phase second band in Fig.~\ref{Fig-PhBand}.

\begin{figure}
\includegraphics[width=\columnwidth]{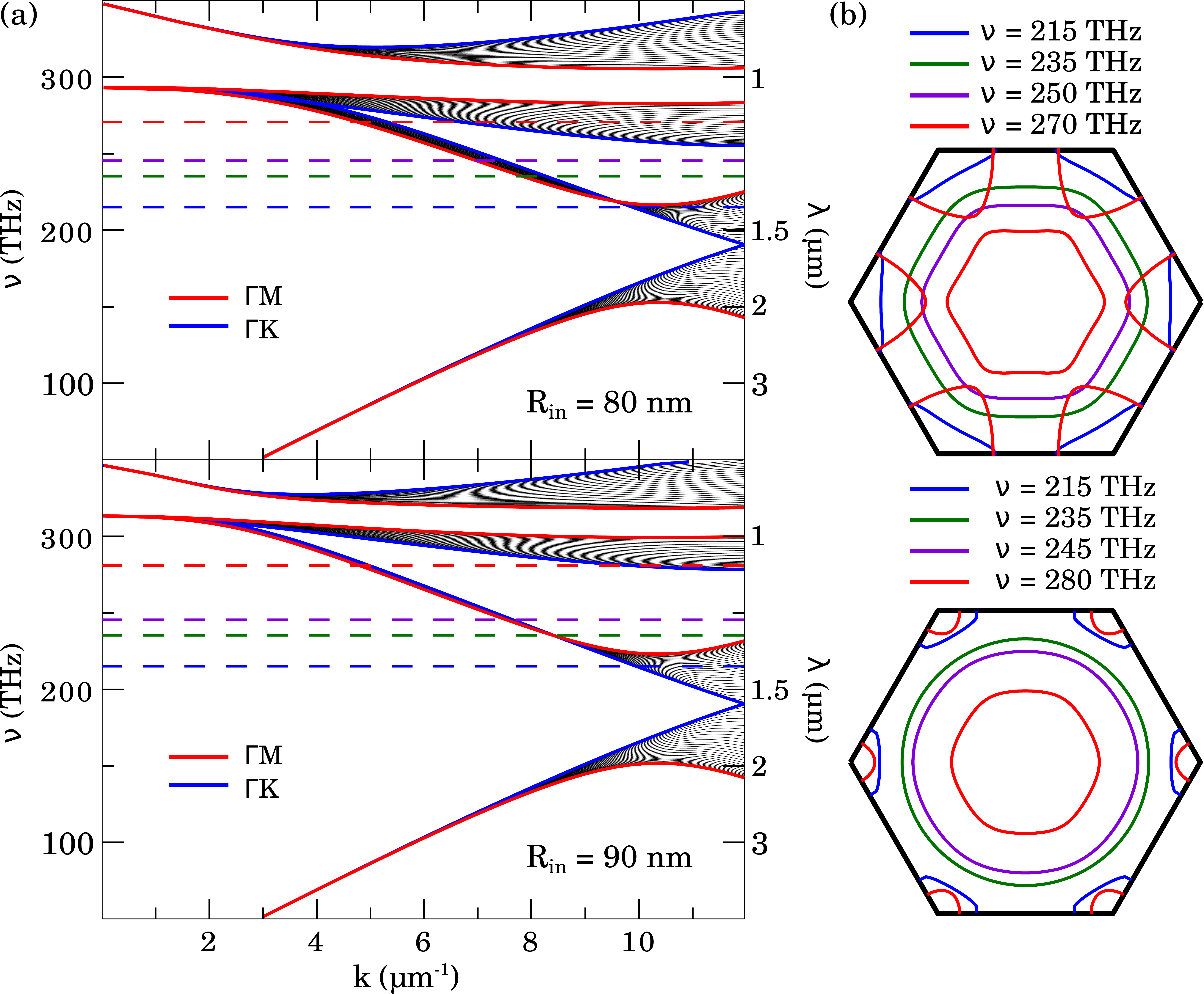}
\caption{(Color online) (a) Photonic band structure as in Fig.~\protect{\ref{Fig-PhBand}} ($a=$ 350 nm, $R_{\mathrm{out}}=$ 170 nm), 
but superimposing various (30) bands at different directions within the first Brillouin zone in between 
$\Gamma$M (red curves) and $\Gamma$K (blue curves). 
(b) Equifrequency surfaces for several frequencies (dashed horizontal lines in (a)) covering mainly the second band. 
In both (a,b): (top)  $R_{\mathrm{in}}=$ 80 nm and (bottom) $R_{\mathrm{in}}=$ 90 nm.}
\label{Fig-APhBand+Isofreq}
\end{figure}

To this end, we represent in Fig. ~\ref{Fig-APhBand+Isofreq}(a) the lowest four photonic bands, upon superimposing the curves 
at various propagation directions in between the maximum symmetry directions, $\Gamma$M and $\Gamma$K (red and blue curves, 
respectively), covering all relevant directions within the first Brillouin zone. In this manner, the partial gaps and the full gap at 300 THz are 
clearly observed.  More importantly, the  nearly isotropic character of the second band is inferred, improving upon slightly increasing the 
NW inner radius from a dispersion of 10 THz ($\sim 5\%$) with wavevector direction shift for $R_{\mathrm{in}}=$ 80 nm,  to a mere 2 THz 
($\sim 1\%$) for $R_{\mathrm{in}}=$ 90 nm. 

In order to clarify isotropy of the wave propagation properties in the second band we plot in Fig.~\ref{Fig-APhBand+Isofreq}(b) 
its equifrequency surfaces (EFSs) for several frequencies [horizontal lines in Fig.~\ref{Fig-APhBand+Isofreq}(a)], 
bearing in mind the hexagonal symmetry of the crystal. 
In the beginning of the second band, at 215 THz ($1.4\,\mu$m), we have the well known pockets due to the partial gap along the 
$\Gamma$M direction, whereas the only allowed propagation modes  are close to the $\Gamma$K directions. 
For frequencies above the partial gap [see Fig.~\ref{Fig-APhBand+Isofreq}(a)], all the mode propagation directions 
are again available and the EFSs become more circular, as shown at 235 THz ($1.28 \mu$m): Isotropy is indeed remarkable for the NW 
$R_{\mathrm{in}}=$ 90 nm inner radius at 245 THz (1.22 $\mu$m), with a nearly circular EFS. Then, as the frequency is further increased,  
at 270 THz (1.11 $\mu$m) and at 280 THz (1.07 $\mu$m), the EFSs of the second bands shrink and deform (become more hexagonal), 
overlapping with the third-band pockets appearing along the $\Gamma$K directions, thus breaking  isotropy. 
Overall, it should be emphasized that there is a wide frequency range (215--260 THz) in which the EFSs are almost circular 
(specially for $R_{\mathrm{in}}=$ 90 nm) and the crystal might be described as an isotropic negative-index medium.
\begin{figure}
\includegraphics[width=\columnwidth]{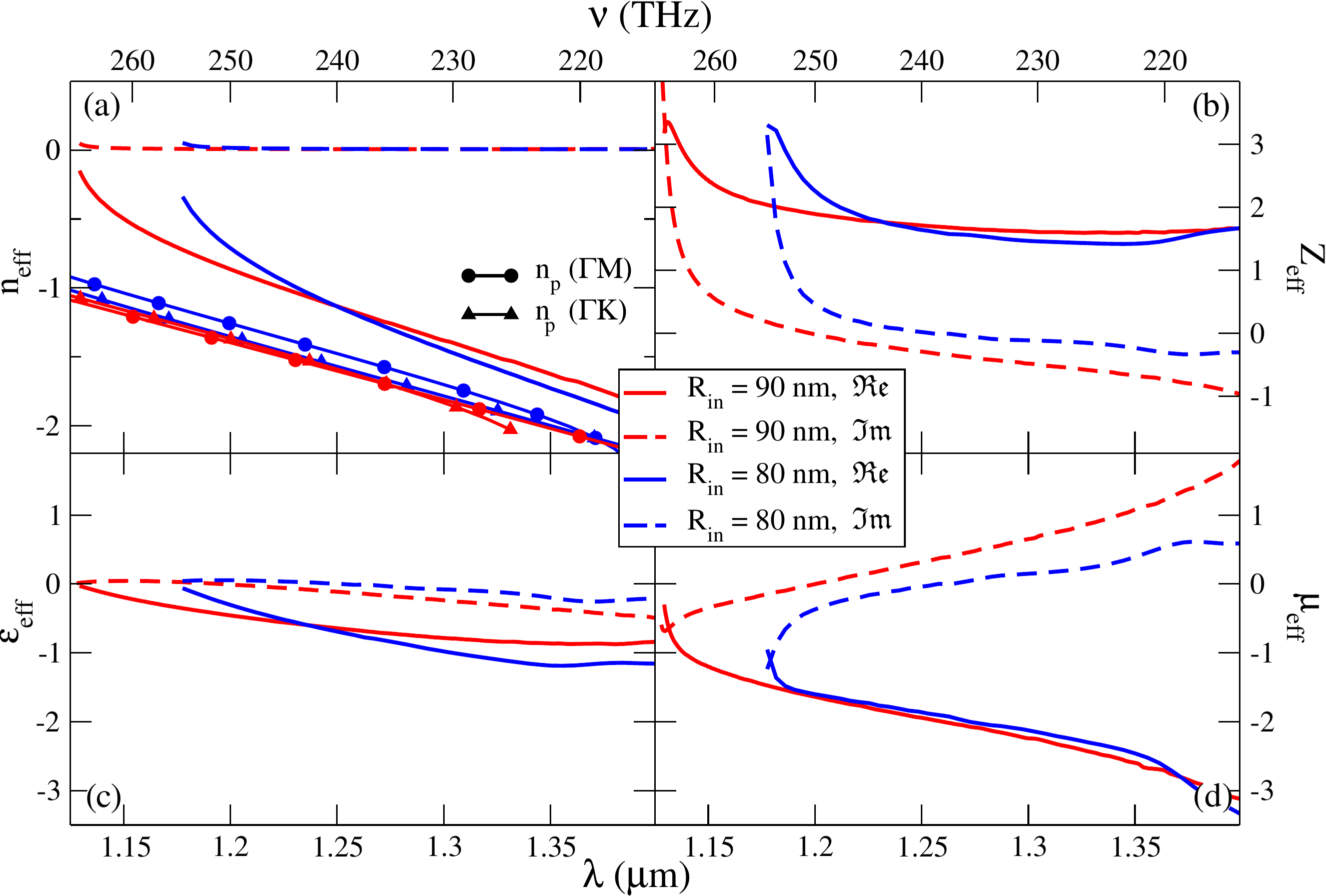}
\caption{(Color online) Spectral dependence (within the second NIM band) of the refractive index $n_{\mathrm{eff}}$ (a), impedance  $Z_{\mathrm{eff}}$ (b), 
effective dielectric permittivity $\varepsilon_{\mathrm{eff}}$ (c) and magnetic permeability $\mu_{\mathrm{eff}}$ (d) 	retrieved from the 
S-parameters for Ag@Si core-shell nanowires, arranged in a hexagonal lattice with lattice parameter $a=$ 350 nm, with outer radius 
$R_{\mathrm{out}}=$ 170 nm, and two inner radii: $R_{\mathrm{in}}=$ 80 nm (blue lines) and $R_{\mathrm{in}}=$ 90 nm (red 
lines). Solid curves: real parts; dashed curves: imaginary parts. The phase index $n_p$ obtained from the photonic bands in 
Fig.~\protect{\ref{Fig-PhBand}} is also plotted (with symbols, top left).}
\label{Fig-effconst}
\end{figure}

In the light of these results, it seems possible to make a fully isotropic 2D NW photonic crystal with negative index of refraction by taking 
advantage of both the negative index of the second band, due to the strong electromagnetic resonances of the isolated NWs, and of the 
isotropy that the photonic NW arrays shows within such band. To this aim, we first show in Fig.~\ref{Fig-effconst} the retrieved effective 
parameters (S-parameter retrieval procedure \cite{Chen2004,Smith2005}, through FEM calculations as detailed in the appendix) 
for the electric permittivity and magnetic permeability, along with the resulting index of refraction and impedance: The 
doubly negative behavior with $\Re(\varepsilon_{\mathrm{eff}}),\Re(\mu_{\mathrm{eff}})<0$ within the NIM band  is evident 
in Fig.~\ref{Fig-effconst}. Low losses are also obtained: $\Im(\varepsilon_{\mathrm{eff}}),\Im(\mu_{\mathrm{eff}})\sim 0$ 
[note that both might be slightly negative,  in principle considered unphysical, but in practice as a result of the limitations of the parameter 
retrieval procedure with $\Im(n_{\mathrm{eff}})\sim 0$]. The effective index is negative within the NIM band with nearly negligible losses:  
$\Re(n_{\mathrm{eff}})<0,\Im(n_{\mathrm{eff}})\sim 0$.  Indeed, the retrieved refractive index  for $R_{\mathrm{in}}=$ 90 nm yields an 
extremely large value of the $f.o.m.\sim 200$. Recall that the absorptive losses of a single core-shell NW are very small \cite{SR2012}, so 
that the remaining (small, but non-negligible) NIM losses are very likely due to scattering. The phase index $n_p$ obtained from the 
photonic bands in Fig.~\ref{Fig-PhBand} is also plotted, in good agreement with $\Re(n_{\mathrm{eff}})$.
\begin{figure}
\includegraphics[width=\columnwidth]{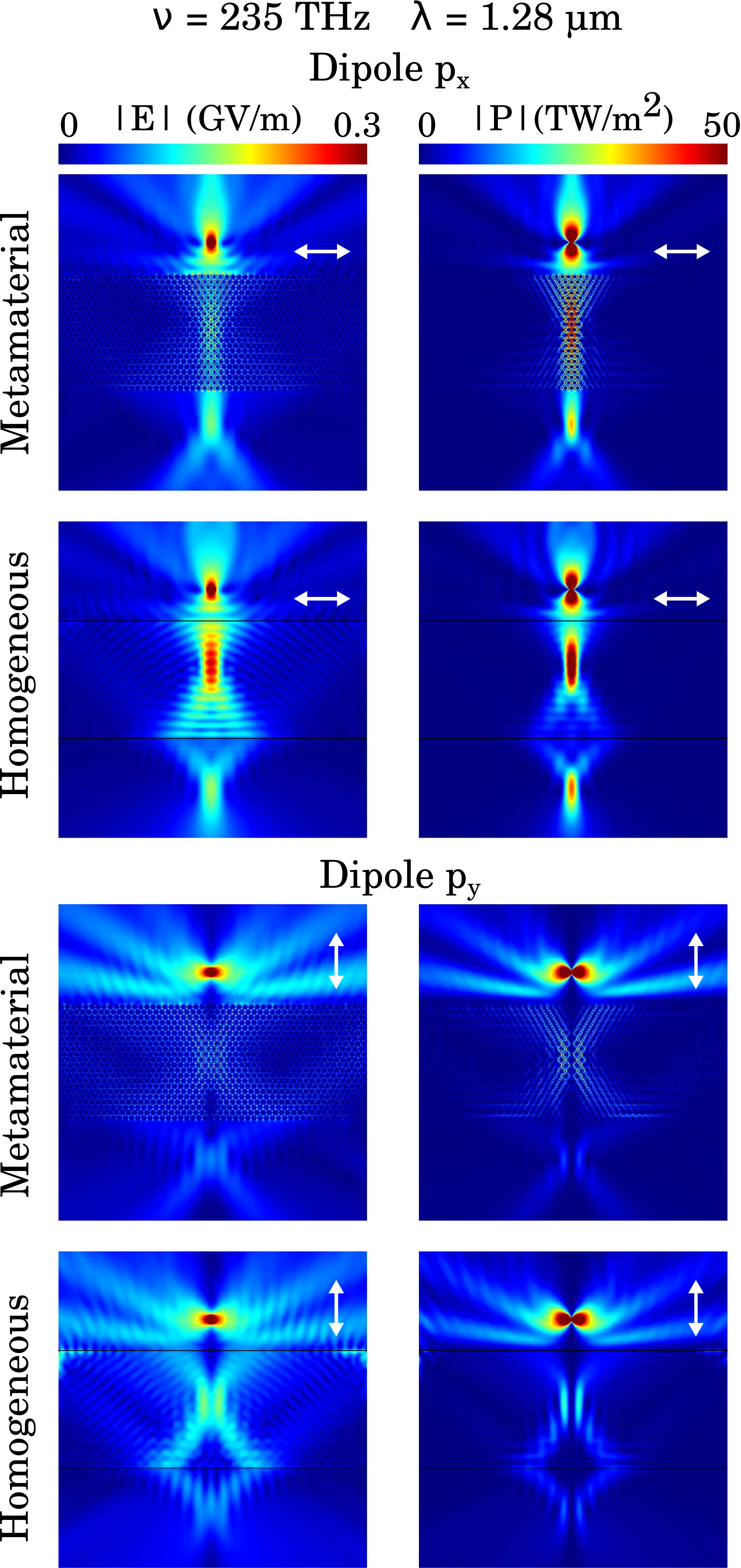}
\caption{(Color online) Transmission of the electromagnetic fields produced by an in-plane point dipole source at 235 THz (1.28 $\mu$m) located at 
1.8 $\mu$m above a slab made of Ag@Si core-shell nanowires with inner and outer radius $R_{\mathrm{in}}=$ 90 nm and 
$R_{\mathrm{out}}=$ 170 nm, respectively, arranged in a hexagonal lattice with lattice parameter $a=$ 350 nm: Slab surface along the 
$\Gamma$M direction. Left: Norm of the in-plane electric field for a dipole parallel (top, $p_{x}$) and perpendicular (bottom, $p_{y}$) to the 
surface slab respectively. Right: Norm of the Poynting vector for a dipole parallel (top) and perpendicular (bottom) to the surface slab, 
respectively. The results for a homogeneous slab are also shown, with $\varepsilon_{\mathrm{eff}}=-0.734-\imath 0.157$ and 
$\mu_{\mathrm{eff}}=-2.094+\imath 0.532$, so that $n_{\mathrm{eff}}=-1.264+\imath 0.0069$ ($f.o.m.\sim 180$) as obtained in  
Fig.~\protect{\ref{Fig-effconst}} at 1.28 $\mu$m.}
\label{Fig-Slab}
\end{figure}

Incidentally, note that $\varepsilon_{\mathrm{eff}}$ and $\mu_{\mathrm{eff}}$ do not become -1 at the same 
frequency, which results in an impedance $Z_{\mathrm{eff}}\equiv(\mu_{\mathrm{eff}}/\varepsilon_{\mathrm{eff}})^{1/2}\gtrsim 1$. Further 
tuning of the core-shell NW parameters could be done in order to achieve the ideal $Z_{\mathrm{eff}}=1$; despite that, the resulting 
impedance is only slightly larger than 1 so that reflections at the interface are small, as will be shown below. However, at this point the 
important issue needs to be addressed whether or not this structure behaves as an isotropic homogeneous medium.

\section{Flat-lens focusing: Isotropic NIM behavior}

To verify that the core-shell NW array presents an isotropic negative index of refraction, rather than considering plane wave negative 
refraction (which has been confirmed up to large angles of incidence in Ref.~\onlinecite{SR2012}), we study the imaging of a point 
source through a slab, which no doubt places a more stringent test for it involves all plane wave angular components.
We carry out numerical simulations of the transmission of the electromagnetic field radiated by an in-plane point dipole source 
at 235 THz (1.28 $\mu$m) placed in front of a finite (22 NW thick, namely, 6.7 $\mu$m long) slab made of the structure studied 
in the previous section (see the appendix for the parameters of the FEM calculations); we focus on the 
$R_{\mathrm{in}}=90$ nm case, which exhibits slightly higher isotropy. In Fig.~\ref{Fig-Slab} we plot the norm of the electric field 
(left column) and the Poynting vector (right column), respectively, for a dipole parallel to the surface  of the slab (top), clearly revealing the 
effects of a medium that possesses a negative index of refraction. At the slab interfaces, light undergoes negative refraction and its field 
components are focused twice, one inside and other outside of the slab: The  image produced on the other  side of the slab presents 
similarities with the real one. For a dipole perpendicular to the slab (see Fig.~\ref{Fig-Slab}, bottom), again, we see the same 
phenomenology of negative refraction through the slab with the expected focusing effect for this dipole orientation. In addition, it is 
remarkable the fact that the Poynting vector is always zero along the propagation direction parallel to the dipole moment axis, so that light 
undergoes no significant spurious scattering losses when traveling through the photonic crystal; which in turn contributes to the effective 
medium parameters as losses, reportedly extremely small in this NIM \cite{SR2012}, and thus in agreement with the above simulations. 
\begin{figure}
\includegraphics[width=\columnwidth]{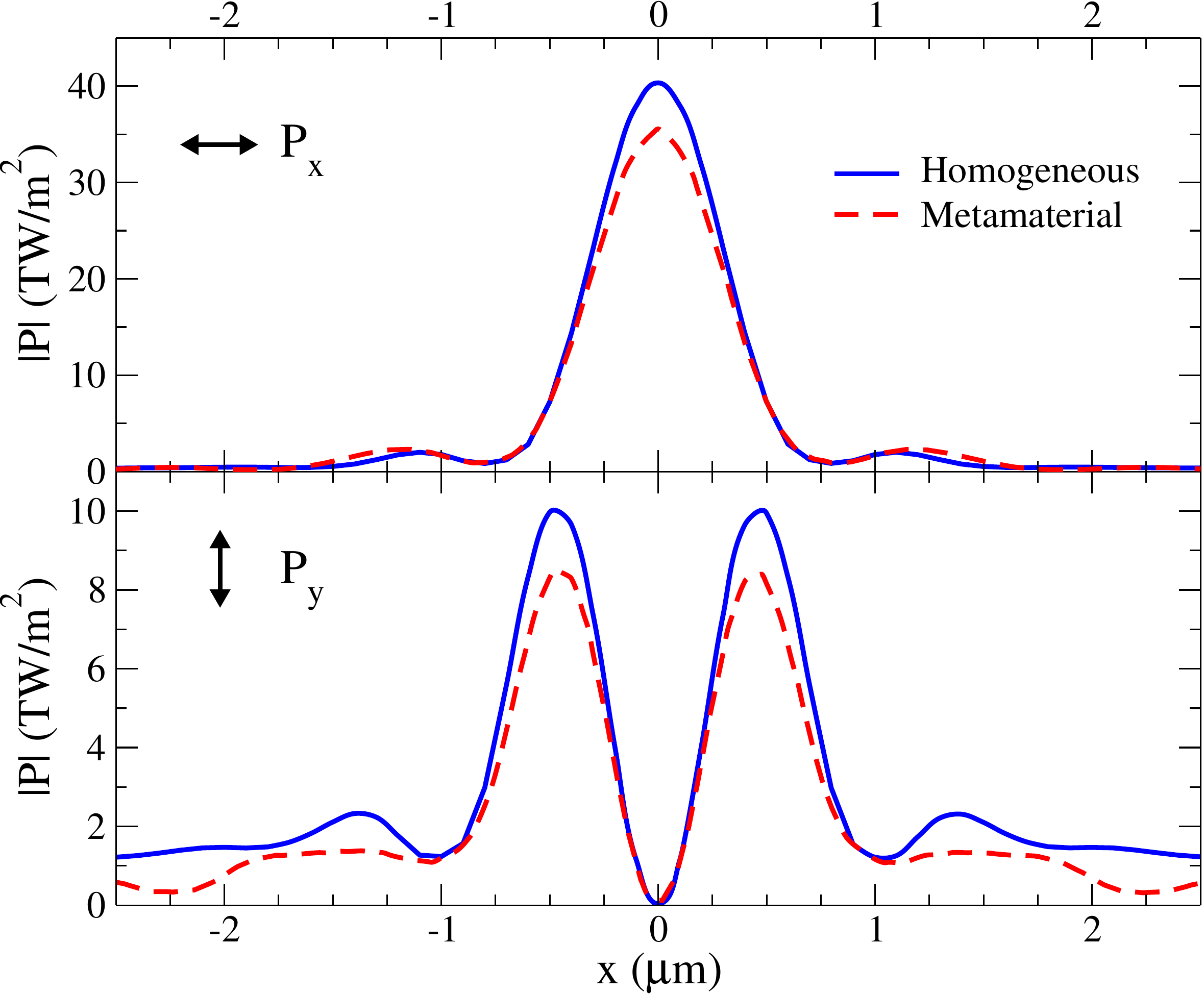}
\caption{(Color online) Poynting vector norms from Fig.~\protect{\ref{Fig-Slab}} at the image plane.}
\label{Fig-Focus}
\end{figure}

Therefore the main features of flat lensing are fully retrieved \cite{Pendry2000,Cubukcu2003,Grbic2004,Xu2013}.  
Furthermore, the agreement is remarkable with the results (shown also in  Fig.~\ref{Fig-Slab}) obtained upon considering 
an isotropic and homogeneous slab with effective parameters at 1.28 $\mu$m given in Fig.~\ref{Fig-effconst}: 
it should be emphasized that the (linear) color scale of both the electric field (left) and Poynting vector (right) norms are 
identical in all plots (columns), either for the NW array or for the homogeneous medium, and also that the scale is neither saturated 
nor altered within the slabs. Remarkably, note that very similar transmitted foci are produced: this is explicitly shown in 
Fig.~\ref{Fig-Focus}, where the power transmitted at the image plane is presented for both the NW array and the homogeneous lenses. 
The full-width at half-maximum is of the order of 0.65 $\mu$m$\simeq\lambda/2$, yielding subwavelength resolution.  
Of course, since the source is not located deep in the near-field, superlensing in the proper sense \cite{Pendry2000} cannot be observed 
and resolution is diffraction limited (ultimately, resolution should be limited for the NW array to NW diameter). 
Overall, we clearly observe the phenomenology of (all-angle) negative refraction at the interface of the slab and the flat-lens focusing 
process. This  supports the highly isotropic, low-loss NIM behavior of the core-shell NW array, and the accuracy 
of the retrieved parameters, in turn ruling out photonic crystal collimation effects 
\cite{Foteinopoulou2003a,Garcia-Pomar2005,Decoopman2006}. 
\begin{figure}
\includegraphics[width=\columnwidth]{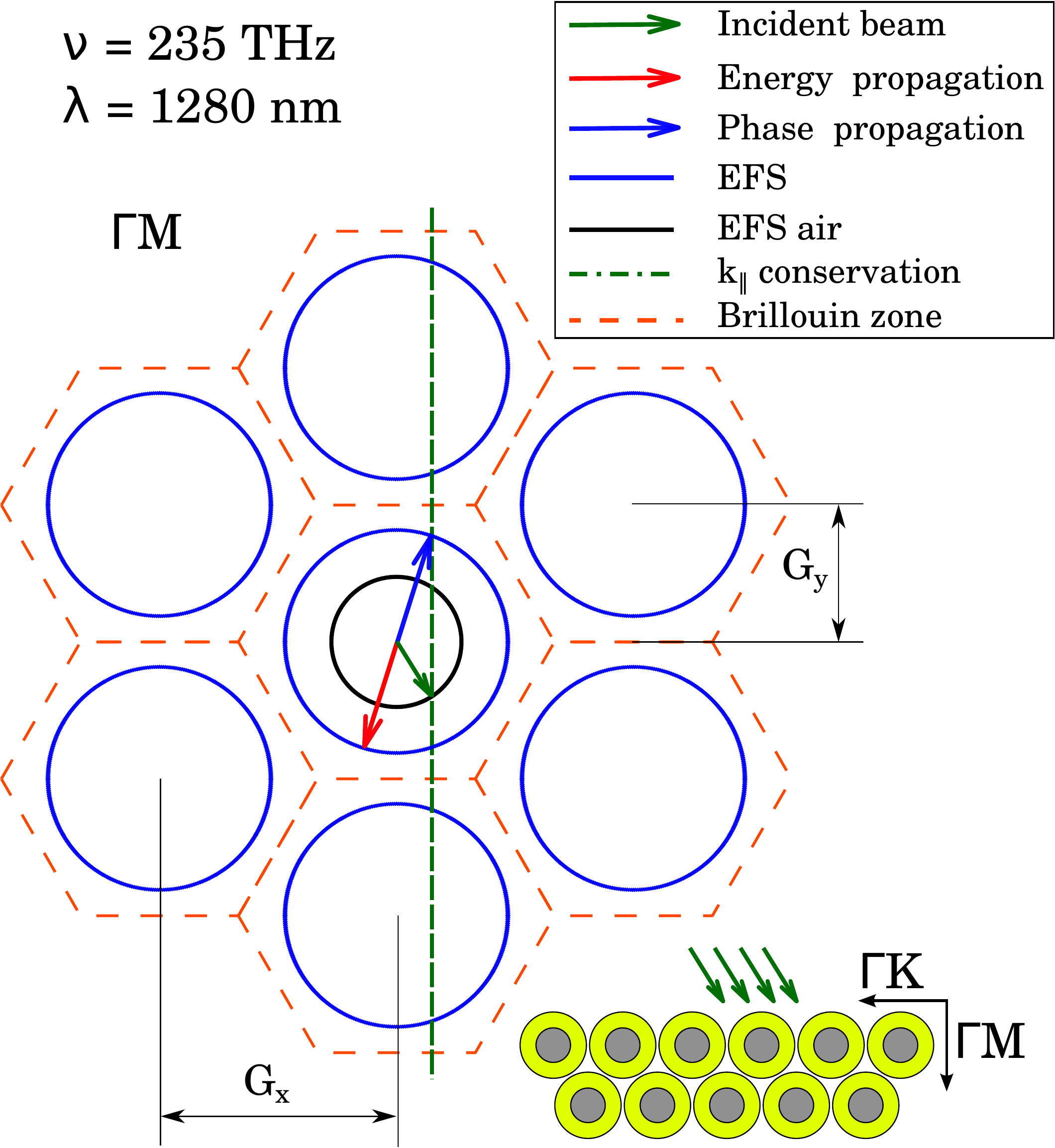}
\caption{(Color online) EFS at 235 THz (1.28 $\mu$m) (blue circle) of a  hexagonal lattice (lattice parameter $a=$ 350 nm, Brillouin zones 
delimited by red dashed  hexagons) of Ag@Si core-shell nanowire  with inner and outer radius $R_{\mathrm{in}}=$ 90 nm and 
$R_{\mathrm{out}}=$ 170 nm. EFS of light impinging from vacuum/air is shown as a black circle. A dashed green (vertical) line denotes 
wavevector conservation along the  $\Gamma$M direction for a particular angle of incidence, refracted accordingly as indicated by 
phase/energy (blue/red, respectively) vectors.}
\label{Fig-EFS}
\end{figure}

Nonetheless, we plot in Fig.~\ref{Fig-EFS} the EFSs for the particular case shown in Fig.~\ref{Fig-Slab}. 
Momentum conservation along the impinging surface (wavevector component along $\Gamma$M) is explicitly denoted by the green 
dashed line for a particular angle of incidence. Such plane wave component should be properly refracted with negative phase velocity (blue 
vector) and positive Poynting (red) vector. Since the EFS is nearly circular and larger than the air light cone, even plane wave components 
impinging from vacuum at grazing angles are negatively refracted. Note also that no additional diffracted  components are expected. 
Therefore, all components of the dipole source should undergo negative refraction in a consistent manner; this corroborates the above 
presented dipole focusing through a slab (see Fig.~\ref{Fig-Slab}).
\begin{figure}
\includegraphics[width=\columnwidth]{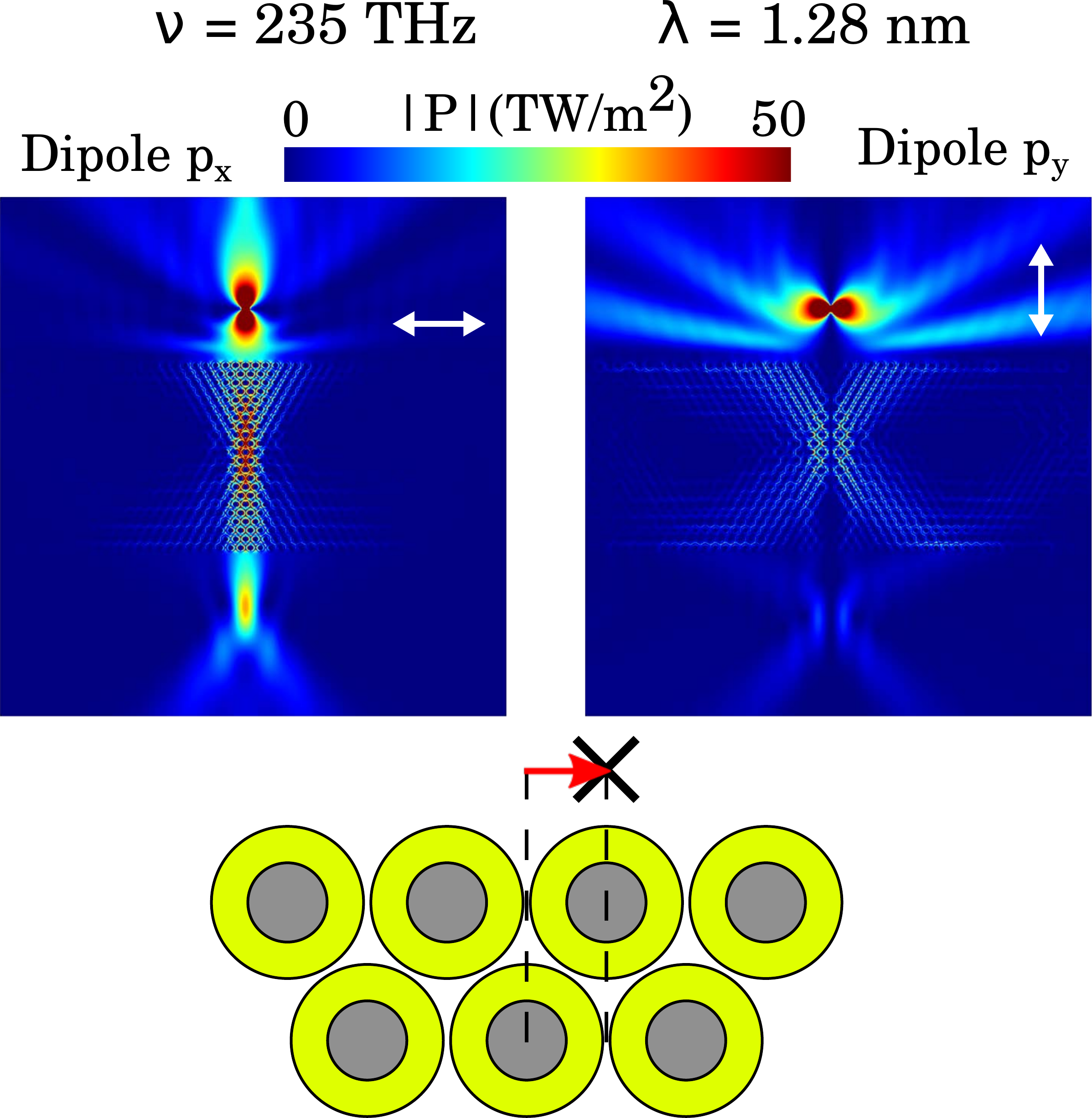}
\caption{(Color online) Top: Transmission through a slab made of a hexagonal array of Ag@Si core-shell nanowires ($a=$ 350 nm, 
$R_{\mathrm{in}}=$ 90 nm, $R_{\mathrm{out}}=$ 170 nm, $\Gamma$M surface direction), of the electromagnetic fields produced 
by an in-plane point dipole source at 235 THz (1.28 $\mu$m), as in Fig.~\protect{\ref{Fig-Slab}} (Poynting vector only), but for 
vertical/horizontal dipoles shifted half-period parallel to the slab surface as depicted (bottom). 
The results for the equivalent homogeneous slab are identical to those in Fig.~\protect{\ref{Fig-Slab}}, and thus not 
shown.
}
\label{Fig-Dip-shiftx}
\end{figure}

\begin{figure}
\includegraphics[width=\columnwidth]{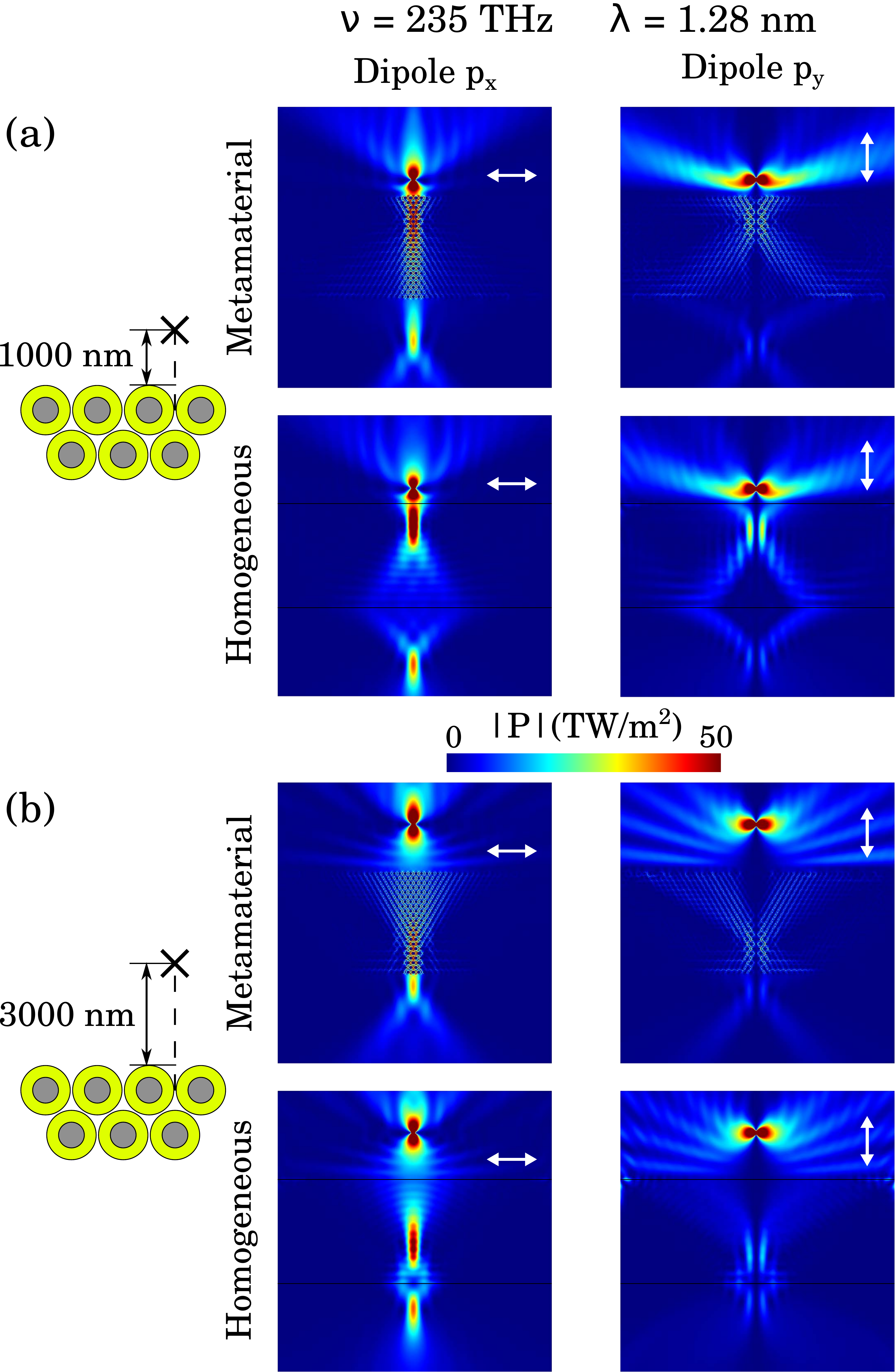}
\caption{(Color online) Transmission through a slab made of a hexagonal array of Ag@Si core-shell nanowires ($a=$ 350 nm, 
$R_{\mathrm{in}}=$ 90 nm, $R_{\mathrm{out}}=$ 170 nm, $\Gamma$M surface direction), of the electromagnetic fields produced 
by an in-plane point dipole source at 235 THz (1.28 $\mu$m),  as in Fig.~\protect{\ref{Fig-Slab}} (Poynting vector only), but for 
horizontal (left column) and vertical (right column) dipoles shifted perpendicularly to the slab surface, both down (top) and up (bottom),  
as depicted by the left schemes. Results are also included for a homogeneous slab with 
$\varepsilon_{\mathrm{eff}}=-0.734-\imath 0.157$ and $\mu_{\mathrm{eff}}=-2.094+\imath 0.532$, so that 
$n_{\mathrm{eff}}=-1.264+\imath 0.0069$ ($f.o.m.\sim 180$) as obtained in  Fig.~\protect{\ref{Fig-effconst}} at 1.28$\mu$m.}
\label{Fig-Dip-shiftz}
\end{figure}
Such NIM focusing behavior is in turn preserved upon moving the source. First, we show in Fig.~\ref{Fig-Dip-shiftx} the negligible impact 
of shifting the dipole source (either $p_x$ or $p_y$) by half-period parallel to the surface, as compared to Fig.~\ref{Fig-Slab} and in 
agreement with the homogeneous slab, further ruling out any artifact induced by single meta-atoms. Likewise,
dipole positions are shifted perpendicular to the surface in Fig.~\ref{Fig-Dip-shiftz}, either closer (top) or farther (bottom); since this source 
displacement is more involved than the parallel one above, results are also included for the homogeneous slab. According to the
canonical NIM focusing effect (keeping the slab thickness fixed), if the dipole source is closer (respectively, farther) to the slab, 
the virtual image inside the slab gets closer (respectively, farther), whereas the image behind the slab appears farther (respectively, closer) 
from the slab surface; this is properly reproduced by the numerical calculations for the Ag@Si core-shell nanowire array.

\begin{figure}
\includegraphics[width=\columnwidth]{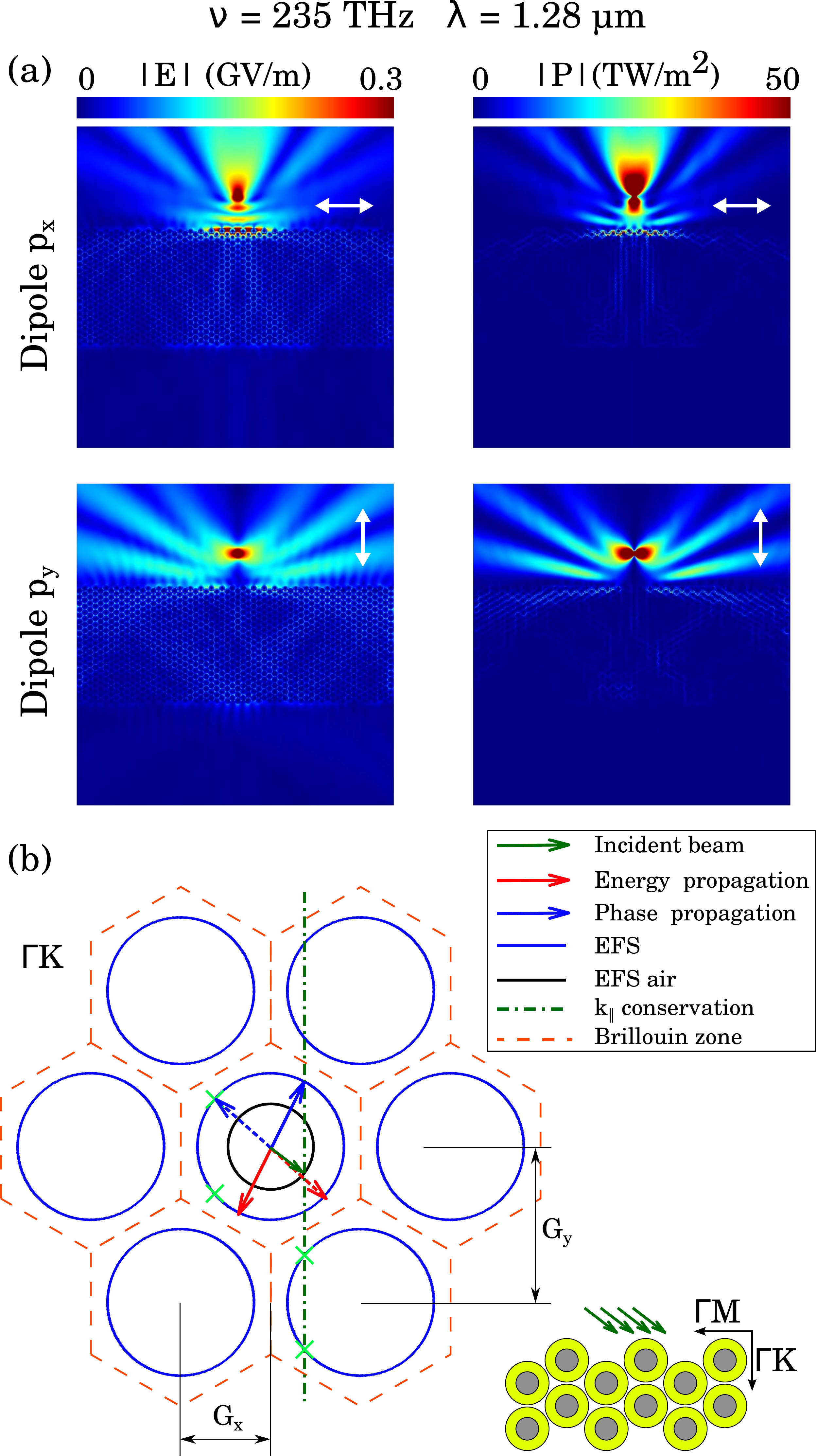}
\caption{(Color online) Top: Transmission through a slab made of a hexagonal array of Ag@Si core-shell nanowires ($a=$ 350 nm, 
$R_{\mathrm{in}}=$ 90 nm, $R_{\mathrm{out}}=$170 nm), of the electromagnetic fields produced by  in-plane point 
(horizontal/vertical) dipole sources at 235 THz (1.28 $\mu$m),  as in Fig.~\protect{\ref{Fig-Slab}}, but for
a slab surface along the $\Gamma$K direction. Bottom: Corresponding EFS (blue curves) of the hexagonal lattice 
along with that of light impinging from vacuum/air (black circle). A dashed green (vertical) line denotes 
wavevector conservation along the  $\Gamma$K direction for a particular angle of incidence, producing 
accordingly two refracted beams as indicated by phase/energy (blue/red, respectively) vectors.
}
\label{Fig-Dip-GK}
\end{figure}
Nonetheless, we would like to stress that the direction along which the array is terminated at the surface plays a crucial role despite 
isotropy. We calculate in Fig.~\ref{Fig-Dip-GK} the dipole source transmission through a slab identical to that used in  Fig.~\ref{Fig-Slab}, 
except for the fact that the  slab surface is along the $\Gamma$K direction: it is evident that NIM focusing is remarkably worse. 
Two effects contribute to hinder it. First, surface roughness is larger along the $\Gamma$K direction, thus leading to larger scattering 
losses from single surface nanowires, as observed in Fig.~\ref{Fig-Dip-GK} through both larger electromagnetic fields on the slab surface 
close to the source, and larger reflection losses. Second, momentum conservation parallel to the $\Gamma$K direction leads to 
a scenario different from that along $\Gamma$M: apart from the expected negatively refracted beam, another refracted 
beam appears beyond a relatively small incidence angle as a diffracted beam at nearest Brillouin zone. This implies that dipole source 
components at a certain angle with respect to the surface normal exhibit two diffracted beams that preclude proper NIM focusing: such 
behavior is evidenced in Fig.~\ref{Fig-Dip-GK} by crossing beams inside the slab.


\section{Refraction through a prism: Tolerance to disorder}

\begin{figure}
\includegraphics[width=\columnwidth]{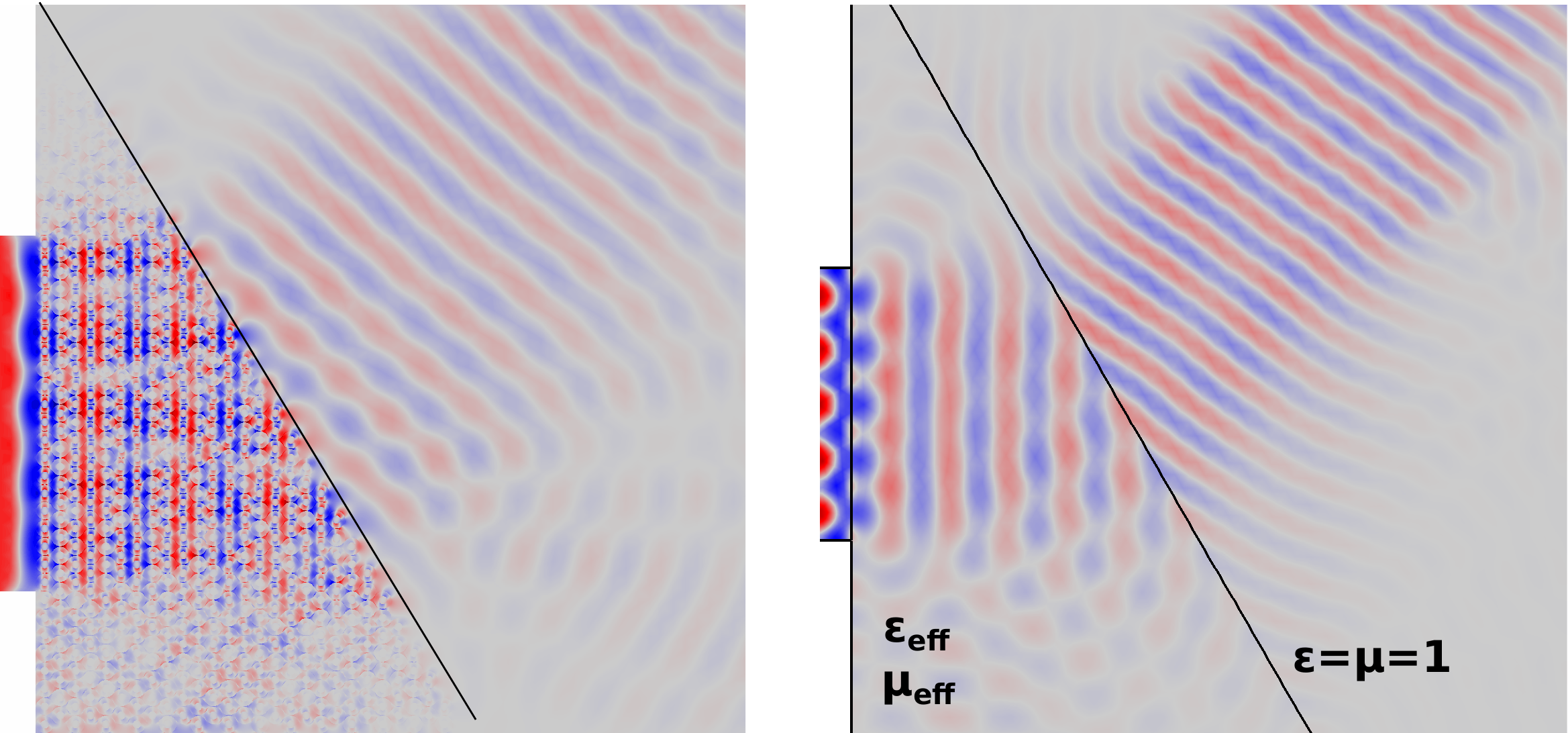}
\caption{(Color online) (a) Refraction  through a $\theta=30^{\circ}$ prism made of a NIM built with Ag@Si core-shell NWs with 
$R_{\mathrm{in}}=80$ nm 
and $R_{\mathrm{out}}=170$ nm arranged in a hexagonal lattice ($a=$ 350 nm), illuminated by a transverse electric (truncated) plane 
wave (1.215 $\mu$m) parallel to the left interface:  contour maps of the out-of-plane, $z$-component of the magnetic field amplitude. (b) 
Same as in (a), but for a homogeneous medium with $\varepsilon_{\mathrm{eff}}=-0.45+\imath 0.04$ 
and $\mu_{\mathrm{eff}}=-1.7-\imath 0.21$, so that $n_{\mathrm{eff}}=-0.88+\imath 0.01$ ($f.o.m.\sim 80$)
as obtained in Fig.~\protect{\ref{Fig-effconst}}.
}
\label{Fig-Prism}
\end{figure}
In the preceding section we have demonstrated by means of flat-lens focusing  that this doubly-resonant NW 
array behaves effectively as an actual isotropic (bulk) optical  NIM when arranged in a close hexagonal lattice packet, 
stemming from the isotropic negative-phase behavior of the second photonic band.  However, it has been shown that ordered 
configurations that exhibit a negative index of refraction may lose such effective property when the geometry is disordered. 
In this regard, it should be recalled that it has been shown in other cylinder arrays that Mie resonances (rather than Bragg scattering)  
govern in part the appearance of spectrally overlapping, photonic band gaps and higher-order bands \cite{Dominec2014,Rybin2014b}. 
Therefore, if Mie (rather than Bragg) scattering underlies the negative-index behavior, the latter should be more tolerant to disorder than 
usual photonic crystals. 

In order to show the robustness of doubly resonant NWs, let us show first  in Fig.~\ref{Fig-Prism} the classical plane wave refraction 
through a rectangular prism with an angle of $\theta=30^{\circ}$, which actually involves two interfaces terminated in different photonic 
crystal directions (see the appendix for details of the FEM numerical calculations). We can see how the wavefronts leave the prisms above 
the normal, a clear signature of a NIM medium. Indeed, the full simulation of the wave propagation 
(see Video 1 in Ref. \onlinecite{SR2012}, Supplementary Information) revealed that the wavefronts travel backward inside the prism 
(negative phase velocity), characteristic of a system having simultaneously negative permittivity and permeability. 
The homogeneous prism with the retrieved effective parameters nicely reproduces the expected negative refraction, thus  further 
supporting the accuracy of the effective parameters and the NIM behavior of the NW array.

\begin{figure}
\includegraphics[width=\columnwidth]{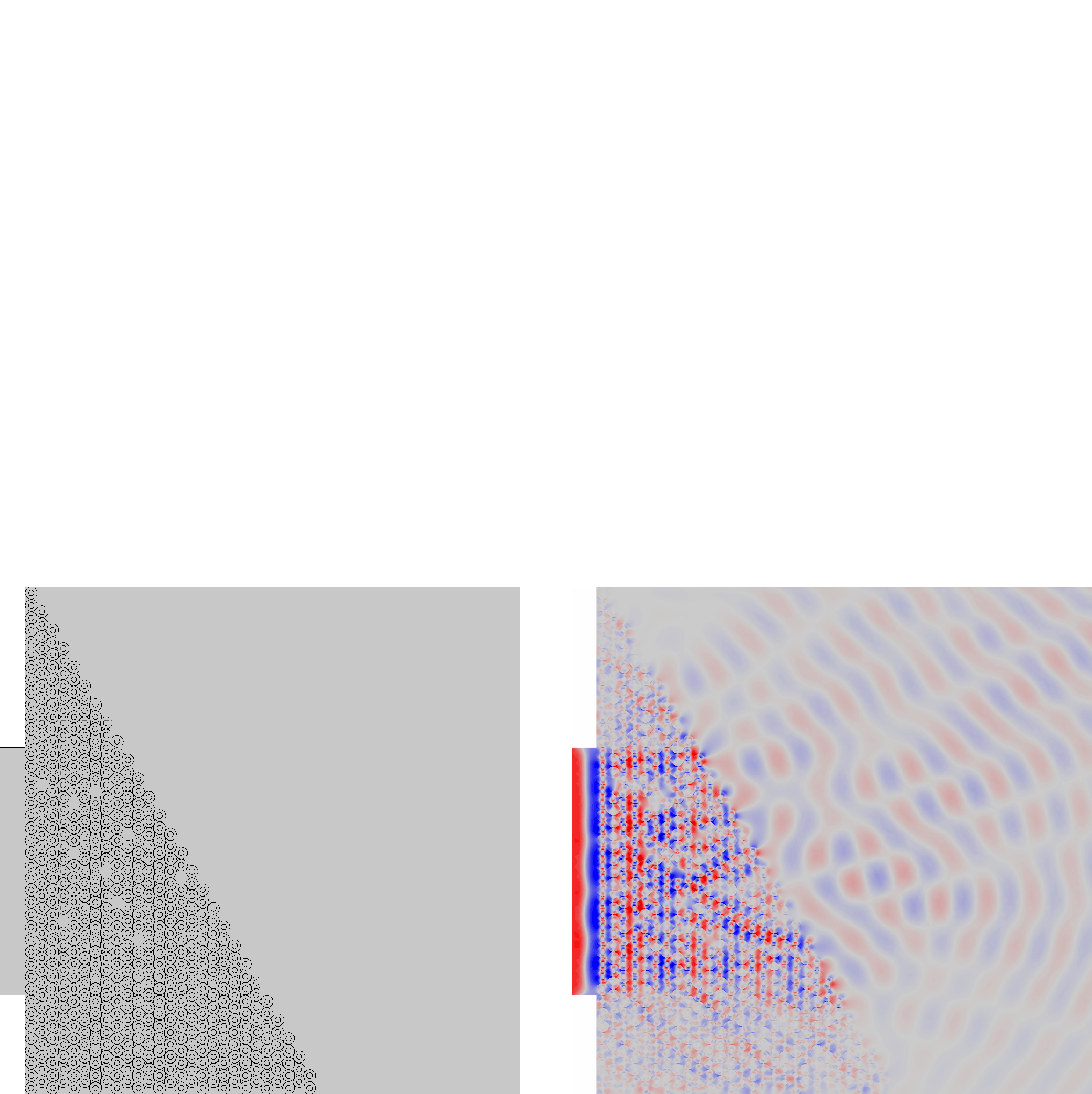}
\caption{(Color online) Negative refraction through disordered  rectangular prisms with an angle of $\theta =30^{\circ}$ built with Ag@Si core-shell 
nanowires with $R_{\mathrm{in}} =$ 80 nm and $R_{\mathrm{out}} =$ 170 nm arranged in a hexagonal lattice ($a=350$ nm) illuminated by 
a (transverse electric) plane wave parallel to the left interface at 1.215 $\mu$m. Disorder is introduced in two different manners: (top left) 
some of the NWs (darker) are shifted by an average of ~1.5 nm; (bottom left) some of the NWs are removed. Right column: Corresponding 
contour maps of the amplitude of the  magnetic field (out-of-plane) component.}
\label{Fig-DisorderPrism}
\end{figure}

We now introduce disorder in the prism geometry used in Fig.~\ref{Fig-Prism}: In essence, some of the constituents are either shifted or 
removed from their original position in the ordered lattice. In Fig.~\ref{Fig-DisorderPrism}(a) we explicitly show the NWs that have been 
shifted from their array positions in the prism and in Fig.~\ref{Fig-DisorderPrism}(b) the plane wave propagation into the disordered prism.  
Although the wavefronts are somewhat perturbed by disorder upon crossing the prism and in the very near field region, we still observe the 
two key feature that characterize the negative index of refraction: an angle of transmission above the normal of the surface and a backward 
propagation of the wavefronts (negative phase velocity).

Moreover, such NIM features are preserved even if stronger scattering disorder is introduced, as shown in the lower panel in 
Fig.~\ref{Fig-DisorderPrism}, where some NWs  have been removed exactly along the wavefront propagation inside the prism. Actually, 
wave distortion is substantially larger than in the preceding case (shifted disorder), clearly observed in the near field close to the outgoing 
prism interface, although the backward propagation inside the prism and the (not so) far-field negatively refracted wavefronts are 
preserved. Bear in mind that, apart from positional disorder, if the NW filling fraction is substantially diminished, a decrease in the (absolute) 
values of the negative effective parameters is expected \cite{SR2012}.  

\begin{figure}
\includegraphics[width=\columnwidth]{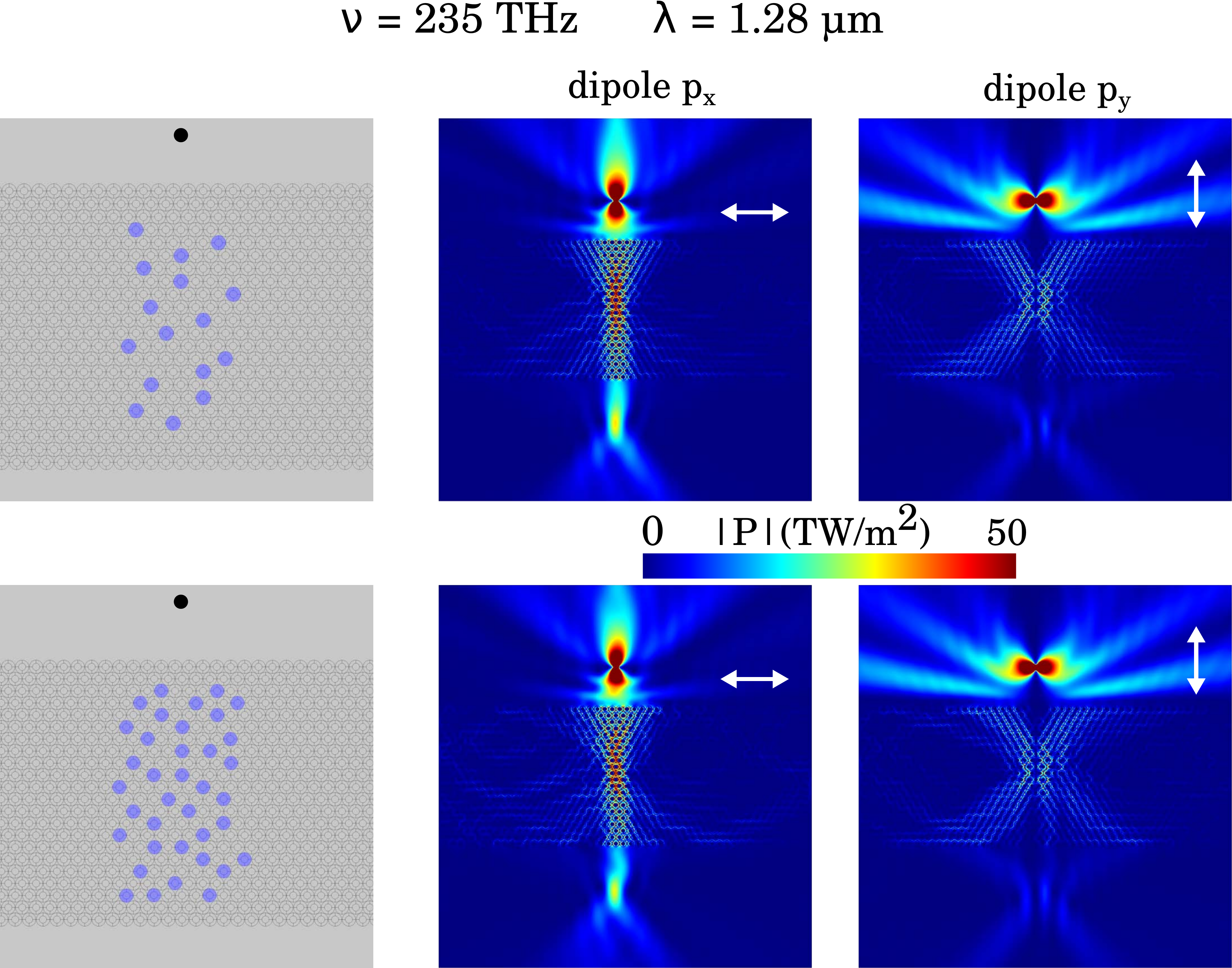}
\caption{(Color online) Transmission through a slab made of a hexagonal array of Ag@Si core-shell nanowires ($a=$ 350 nm, 
$R_{\mathrm{in}}=$ 90 nm, $R_{\mathrm{out}}=$ 170 nm, $\Gamma$M surface direction), of the electromagnetic fields produced 
by  in-plane (horizontal, center, and vertical, right) point dipole sources at 235 THz (1.28 $\mu$m),  as in Fig.~\protect{\ref{Fig-Slab}} 
(Poynting vector only), introducing disorder in the region below the dipoles by shifting the position of the NWs (darker ones) shown in 
the left column schemes: 16 (top row) and 37 (bottom row) NWs are randomly shifted.}
\label{Fig-DisorderSlab}
\end{figure}
Finally, disorder is also introduced in the slab geometry used in Fig.~\ref{Fig-Slab}, again by shifting some of the NWs  from their original 
position in the ordered lattice, as shown in Fig.~\ref{Fig-DisorderSlab}. Note that, to make more relevant the impact of disorder, the shifted 
NWs are located in the central part of the slab, through which most of the dipole transmitted power is concentrated. A moderate amount of 
disorder (top) does not disturb the flat-lens focusing. Only when the number of shifted NWs increases (see Fig.~\ref{Fig-DisorderSlab}, 
bottom), the transmitted pattern is slightly altered, though the transmitted focal spot is still reproduced. Vacancy disorder (not shown here) 
is extremely sensitive in this case to the vacant position for obvious reasons, leading to strong scattering if placed nearby the intermediate 
focus region  inside the slab, as expected. Nevertheless, we have shown that the properties of our closely packed, doubly-resonant 
NWs are not only due to the periodicity of the lattice, but do survive even if substantial (either positional or vacancy-like) disorder is 
introduced, bearing in mind that filling fraction should be preserved. 

\section{Concluding remarks}

In conclusion, we have carried out  a full description of the photonic band structure of doubly-resonant core-shell metallo-dielectric 
nanowire arrays, in order to demonstrate their behavior as low-loss, isotropic bulk optical NIMs, as reported in Ref.\onlinecite{SR2012} 
from the retrieved effective medium parameters. This must be done, as pointed out in Ref.~\onlinecite{Valdivia2012},  to rule out that their 
negative refraction behavior  is mainly a photonic-crystal diffraction effect that depends on the specific geometry sample. The 2D photonic 
band structure has been presented, identifying the second photonic band with negative-index properties. The resulting EFSs in the first 
Brillouin zone are circular for most of this band, thus indicating the full (2D) isotropic character. In addition, such  isotropic NIM behavior is 
very clearly evidenced through numerical simulations in canonical examples for finite geometries: Namely, point dipole transmission 
through a slab and negative refraction through a prism. Both cases are compared to the same configurations, but for a slab/prism 
characterized by homogeneous dielectric permittivity and magnetic permeability exactly given by the effective parameter retrieval (from 
S-parameters), in very good agreement. This further confirms the effective (highly isotropic) bulk NIM behavior, in turn explicitly manifesting 
the extremely low losses and suitable (close to 1) impedance. Finally, disorder is introduced in such finite prisms to support the robustness 
of the NW structures as NIMs, inferring that Mie resonances (rather than Bragg scattering) govern to a large extent their optical properties. 

In this manner, we have rigorously demonstrated (and improved) the highly isotropic and ultra low-loss ($f.o.m.\sim 200$)  character
of Ag@Si nanowire arrays, exhibiting indeed robust flat lensing with subwavelength resolution and dipole position independence, in good 
agreement with the expected behavior for a negative-index homogeneous slab. Our results and procedure serve to fully assess such 
fascinating properties as optical NIMs of the doubly-resonant core-shell metallo-dielectric nanowire arrays, paving the way towards 
realistic fabrication thanks to the tolerance to specific geometric considerations and inhomogeneities.

\appendix*
\section{FEM numerical calculations}

The numerical calculations of photonic band structures and wave propagation through slabs were carried out by means of full 
electromagnetic simulations using a Finite Element Method commercial software (COMSOL Multiphysics 4.4). 
For photonic band structures, an eigenfrequency study is applied to the unit cell upon Bloch boundary conditions 
to reproduce an infinite system. The calculated photonic band direction is determined by the periodicity of the $k$-vector. 

In the case of the metameterial slabs, the simulated space consisted of an hexagonal array of of core-shell nanowires of height 
$H \approx$ 6.7 $\mu$m and length $L \approx$ 35 $\mu$m (22x100 nanowires), representing the slab, centered in a rectangle three times 
thicker than the slab and with the same length. In addition, all the physical system is surrounded by an additional layer of 2 $\mu$m that is 
set as a perfectly matched layer (PML) to absorb all the outgoing radiation. The source of the electromagnetic field is simulated by an in-plane 
electric point dipole. For the homogeneous slab, the simulated space is the same as the metamaterial slab, but replacing the periodic 
array by a rectangle of the same $H$ and $L$, and applying a transition boundary condition to its boundaries, setting its thickness to 1 nm 
and  its optical properties to those of the surrounding medium (air).

In all cases, silicon was assumed lossless in the frequency range studied, with a refractive index of $n_{Si} = 3.5$ and silver material 
constant were taken from experimental values from \cite{Palik1998}. The electromagnetic properties of the homogeneous medium were 
those determined by S-parameter retrieval procedure as in Refs. \cite{Chen2004,Smith2005}, 
and the rest of the space was set to be air ($n_{air} = 1$).

The meshing was done with the program built-in algorithm, which creates a triangular mesh. For photonic band structure calculations, the 
mesh maximum element size (MES) was set to 7.4 nm and the maximum element growth rate (MEGR) to $1.1$. In addition, opposite 
edges of the unit cell were meshed identically by using the Copy Edge feature to ensure the symmetry of the system.  For the 
transmission of the electromagnetic field radiated by a dipole, the MES was set to 100 nm and the MEGR to 1.3.

\begin{acknowledgments}
The authors acknowledge the Spanish "Ministerio de Econom\'{\i}a y Competitividad" for financial support through the
Consolider-Ingenio project EMET (CSD2008-00066), and NANOPLAS+ (FIS2012-31070), MINIELPHO (FIS2012-36113-C03-03)
and FIS2014-55563-REDC.
\end{acknowledgments}


\end{document}